\documentstyle[12pt,epsfig]{article}

\begin{document}

\begin{titlepage}

\rightline{November 2010}

\vskip 2cm

\centerline{\large \bf  
Do magnetic fields prevent mirror particles
}

\vskip 0.4cm
\centerline{
\large \bf 
from entering the galactic disk?}

\

\vskip 1.7cm

\centerline{R. Foot\footnote{
E-mail address: rfoot@unimelb.edu.au}}

\vskip 0.7cm

\centerline{\it School of Physics,}
\centerline{\it University of Melbourne,}
\centerline{\it Victoria 3010 Australia}

\vskip 2cm

\noindent
Recently it has been suggested that magnetic fields prevent mirror particles from
entering the galactic disk, thereby disfavouring the mirror dark matter
explanation of the dark matter direct detection experiments.
We show that mirror particle self interactions will typically randomize the directions of mirror
particles  
on length scales much shorter than their gyroradius. This means that mirror
particles are free to enter the galactic disk and consequently mirror dark
matter remains consistent with all experiments and observations.

\end{titlepage}

Mirror dark matter has emerged as a consistent explanation\cite{m1} of 
the positive signals reported in the DAMA\cite{dama} and CoGeNT\cite{cogent}
dark matter direct detection experiments.
Recall, that this theory assumes the existence of a hidden sector exactly isomorphic 
to the standard model
sector\cite{flv} which interacts with the ordinary particles via 
$U(1)$ kinetic mixing\cite{fh} of strength $\epsilon \sim 10^{-9}$. For a review and more detailed
references to the literature, see e.g. ref.\cite{review}.

In the mirror dark matter theory, the galactic halo consists predominantly of a mirror particle
plasma. The plasma contains the various mirror elements, $e', P', He', O',...$ assumed to be in 
approximate thermal equilibrium due to the self interactions. The condition of hydrostatic 
equilibrium implies that the temperature of the plasma satisfies\cite{fv}:
\begin{eqnarray}
T = {1 \over 2} \bar m v_{rot}^2\ ,
\end{eqnarray}
where $\bar m$ is the mean mass of the particles in the plasma, and $v_{rot} \approx 250$ km/s is
the galactic rotational velocity. Assuming that the plasma is fully ionised and dominated
by $He'$, then $\bar m \approx 4m_p/3$ and $T \approx 0.5$ keV.

The mirror dark matter interpretation\cite{m1} of the DAMA and CoGeNT
direct detection experiments
involves scattering off target nuclei of the dominant mirror metal component of
the halo,
$A'$. The parameters favoured by the data are\cite{m1}:
\begin{eqnarray}
 {m_{A'} \over m_p} &=& 22\pm 8 \nonumber \\
 \epsilon \sqrt{\xi_{A'}} &=& (7 \pm 3) \times 10^{-10}\ ,
 \label{exp}
 \end{eqnarray}
where $\xi_{A'}$ is the halo mass fraction of $A'$.

Recently McDermott {\it et al.}\cite{zurek}, 
using results from Chuzhoy and Kolb\cite{kolb} have argued 
that the galactic magnetic field is strong enough
to prevent mirror particles from entering the galactic disk and any existing 
mirror particles would be expelled by supernova processes.
If valid, this reasoning would strongly disfavour the mirror dark matter
interpretation of the DAMA, CoGeNT and other direct detection experiments.
Let us examine the argument in more detail.
The large-scale magnetic
field in the Milky Way has strength 1-10 $\mu G$ and is mostly parallel to the plane of the galactic
disk. The magnetic field can potentially block charged particles from entering the disk if their 
gyroradius,
\begin{eqnarray}
R_g  = 10^{-9}\ pc \left( {m_{X} \over  m_p} \right) 
\left( {e \over q_X}\right) 
\left( {v_X \over 300\ km/s}\right) \left( {B \over 1 \mu G} \right)^{-1}
\label{b1}
\end{eqnarray}
is larger than the height of the disk (typically 100 pc).
In the above equation, $m_p$ is the proton mass and $v_X, q_X$ and $m_X$ 
are the velocity, charge and mass of the particle $X$.
The argument, as applied by Chuzhoy and Kolb\cite{kolb} to 
charged halo massive particles
seems straightforward.
However, there are two obvious barriers in applying this argument to
mirror particles (as was recently attempted in ref.\cite{zurek}).
Firstly, mirror particle self interactions
mediated by the mirror photon are not negligible and need to be considered. These self 
interactions can
potentially randomize the directions of the mirror particles over distance scales
much less than the gyroradius, $R_g$. Secondly, the mirror particle plasma
is a very complicated medium. The effects of mirror electric and 
magnetic fields could potentially dominate over the very feeble mirror $B$ field induced from
the ordinary galactic magnetic field, which is just $B' = \epsilon B \sim 10^{-15}$ Gauss.
The study of such a plasma would be governed by the laws of magnetohydrodynamics, 
which is quite non-trivial, and we will not
attempt to solve it here.
Instead, we will examine the randomizing effect of mirror particle collisions
which it turns out is sufficient to allow mirror particles to enter the galactic disk in essentially all of the mass range
favoured by the DAMA and CoGeNT experiments.

If we define $\ell$ as the distance scale over which
collisions are able to randomize the direction of the mirror particles, 
then mirror particles
will be able to penetrate the galactic disk so long as the distance $\ell$ is much less than
the gyroradius, $R_g$.
Mirror particle collisions are dominated by elastic Rutherford scattering,
and for $A' - A'$ self interactions in the center of mass frame we have:\footnote{
We use units where $\hbar = c = 1$ throughout.}
\begin{eqnarray}
{d \sigma \over d\Omega} = {\alpha^2 Z'^4  \over 16 m_{A'}^2 v_{cm}^4 \sin^4 {\theta \over 2} } \ ,
\end{eqnarray}
where $v_{cm}$ is the magnitude of the $A'$ velocity in the center of mass frame and $Z'$ is the mirror
atomic number of the $A'$ nucleus.
The cross-section for a single large angle collision, which we define as a 
scattering with $\theta  > \pi/2$, is given by:
\begin{eqnarray}
\sigma^* = { \alpha^2 Z'^4 \pi \over 4 m^2_{A'} v_{cm}^4}\ .
\end{eqnarray}
Of course $A'$ can also undergo multiple small angle collisions.
The minimum angle, $\theta_{min}$,
for which elastic scattering can occur before suppression due to the shielding 
by the charge of neighboring particles in the plasma is given by:
\begin{eqnarray}
\theta_{min} \sim {1 \over \Delta x p_{A'}}
\end{eqnarray}
where $\Delta x \sim 1$ cm is the mean spacing between particles in the plasma at the Earth's location and
$p_{A'} = m_{A'} v $ is the momentum of the particle.
It turns out that scattering is dominated by multiple small 
angle $A'-A'$ collisions which combine to produce a large angular deflection, 
$\theta_{total} \approx \pi/2$, over a distance scale:
\begin{eqnarray}
{\ell (A')} \approx {\pi^2 \over 32\sigma^* n(A') log \left(\Delta x p_{A'}\right)}
\end{eqnarray}
where $n(A')$ is the $A'$ number density.
Evaluating ${\ell (A')}$ at the Earth's location in the galactic disk, we find:
\begin{eqnarray}
{\ell (A') } &\approx & {\pi T^2  \over 2 \alpha^2 Z'^4 
log \left( \Delta x p_{A'}\right)} 
\left[ {m_{A'} \over \zeta_{A'} \rho_{dm}}\right]
\nonumber \\
 &\approx & 1.5 \times 10^{-2}\ pc \ \left[ {T \over 0.5 \ keV}\right]^2
\left[ {10 \over Z'}\right]^4 
 \left[ {10^{-1} \over \zeta_{A'}}\right] \left[ {m_{A'} \over 20m_p}\right] \ , 
\label{a}
\end{eqnarray}
where we have used $T \sim {1 \over 2} m_{A'}\langle v_{cm}^2 \rangle$
and that $n(A') = \zeta_{A'} \rho_{dm}/m_{A'}$.
Here $\rho_{dm} \approx 0.3\ GeV/cm^3$ is the 
local dark matter density\footnote{Note that the data suggest\cite{m1} 
a mirror metal component, $\zeta_{A'} \stackrel{>}{\sim} 
10^{-2}$.}.

The gyroradius for $A'$ is given by:
\begin{eqnarray}
R_g (A') \approx 0.5\ pc \left( {m_{A'} \over 20 m_p} \right)^{1/2} \left( {10^{-9} \over \epsilon}\right)
\left( {10 \over Z'}\right)
\left( {T \over 0.5 \ keV}\right)^{1/2}  
\left( {B \over 1 \mu G} \right)^{-1}
\label{b}
\end{eqnarray}
where $m_p$ is the proton mass and $\epsilon Z' e$ is the electric charge of $A'$ induced by
the photon - mirror photon kinetic mixing of strength $\epsilon$. 
In Eq.(\ref{b}) we have also used the typical velocity $v \sim \sqrt{2T/m} \Rightarrow 
v \sim \sqrt{T/0.5keV} \sqrt{m_p/m_{A'}}\  300\  km/s$.
Comparing Eqs.(\ref{a},\ref{b}) we find that:
\begin{eqnarray}
{\ell (A') \over R_g (A') } \approx 0.03 \left[ {T \over 0.5 \ keV} \right]^{3/2} 
\left[ {10 \over Z'}\right]^3
\left[ {m_{A'} \over 20m_p}\right]^{1/2} \left[ {\epsilon \over 10^{-9}}\right] 
\left[ {10^{-1} \over \zeta_{A'}}\right]\left[{B \over 1 \mu G}\right]\ .
\label{dd}
\end{eqnarray}
Thus we find that $\ell (A')/R_g (A') \ll 1$ for $Z' \stackrel{>}{\sim} 8$ 
[$\Rightarrow m_{A'}/m_p \stackrel{>}{\sim} 16$].
We therefore do not expect the magnetic field to prevent
$A'$ particles from entering the galactic disk for essentially all of the mass range favoured by
the experiments, Eq.(\ref{exp}).

Observe that the values for $\ell/R_g$ for $He'$ and $e'$ are given by
$\ell (He')/R_g (He') \sim 10^{-1}$ and $\ell (e')/R_g (e') \sim 10^2$. 
We do not, however, expect a situation where $A'$ (and $He'$) particles can enter the disk
and the $e'$ component cannot. Put simply,
the electrical forces induced by any bulk charge
separation would quickly overwhelm the feeble $\epsilon B$ force, and
such forces would act most effectively on the $e'$ component due to its
relatively large $charge/mass$ ratio.
These mirror electrical forces would act  
to keep the mirror plasma approximately neutral
(over scales such as $\sim R_g^3$)
and thus it does not matter if $\ell (e')/R_g (e') > 1$. 
The $A'$ can enter the galactic 
disk and mirror electrical forces will ensure that the $e'$ component follows  
ensuring approximate mirror charge
neutrality of the plasma. 

From Eq.(\ref{dd}), mirror nuclei lighter than about $m_{O'} = 16m_p$ might have $\ell \sim R_g$
if the magnetic field is relatively large $B \approx 10 \mu G$.
Thus, it might be possible that the galactic magnetic field at least partially excludes 
such light mirror nuclei. 
However, as mentioned earlier, the mirror particle plasma is a complicated medium. It is certainly
conceivable that
the effects of the mirror electric and magnetic fields induced within the plasma
will dominate over the tiny mirror B field ($\sim 10^{-15}$ Gauss) induced from the galactic magnetic field
and thus no firm conclusions can be made regarding the exclusion of the light mirror nuclei from
the galactic disk.

In conclusion, we have examined the question of whether magnetic fields will
prevent mirror particles from entering the galactic disk.
We have shown that mirror particle self interactions will typically randomize the directions of mirror
particles on length scales much shorter than their gyroradius. 
This means that mirror
particles 
are free to enter the galactic disk. Consequently mirror dark
matter remains a consistent explanation of the DAMA, CoGeNT and other direct detection experiments.

\vskip 1cm

\noindent 
{\bf Acknowledgments}
\vskip 0.2cm
\noindent
This work was supported by the Australian Research Council.


\begin{thebibliography}{999}


\bibitem{m1}

R. Foot, Phys. Lett. B692, 65 (2010) [arXiv: 1004.1424];
Phys. Rev. D82: 095001 (2010) [arXiv: 1008.0685] and 
references there-in.



\bibitem{dama}
R. Bernabei {\it et al.} (DAMA Collaboration), 
Eur. Phys. J. C56, 333 (2008) [arXiv: 0804.2741]; arXiv: 1002.1028;
Riv. Nuovo Cimento. 26, 1 (2003)
[astro-ph/0307403]; Int. J. Mod. Phys. D13, 2127 (2004); Phys.
Lett. B480, 23 (2000).



\bibitem{cogent}
C. E. Aalseth {\it et al.} (CoGeNT Collaboration), arXiv: 1002.4703.


\bibitem{flv}
R. Foot, H. Lew and R. R. Volkas, Phys. Lett. B272, 67 (1991);
Mod. Phys. Lett. A7, 2567 (1992).



\bibitem{fh}
R. Foot and X-G. He, Phys. Lett. B267, 509 (1991).


\bibitem{review}
R. Foot, Int. J. Mod. Phys. D13, 2161 (2004) [astro-ph/0407623]; 
Int. J. Mod. Phys. A19, 3807 (2004) [astro-ph/0309330].



\bibitem{fv}
R. Foot and R. R. Volkas,
Phys. Rev. D70, 123508 (2004) [astro-ph/0407522].



\bibitem{zurek}
S. D. McDermott, H-B. Yu and K. M. Zurek,
arXiv: 1011.2907 \ v1.



\bibitem{kolb}
L. Chuzhoy and E. W. Kolb, 
JCAP 0907, 014 (2009) [arXiv: 0809.0436].



\end{thebibliography}
\end{document}